\documentclass[aps,pre,showpacs,twocolumn,groupedaddress]{revtex4-1}

\usepackage{graphicx}
\usepackage{dcolumn}
\usepackage{bm}
\usepackage{enumerate}
\usepackage{amsmath}

\begin{document}


\title{ \bf The Kuramoto Model with Time-Varying Parameters}


\author{Spase Petkoski}

\author{Aneta Stefanovska}
\email[]{aneta@lancaster.ac.uk}
\affiliation{Department of Physics, Lancaster University, Lancaster LA1 4YB, United Kingdom}


\date{\today}

\begin{abstract}

We analyze the Kuramoto model generalized by explicit consideration
of deterministically time-varying parameters. The oscillators' natural
frequencies and/or couplings are influenced by external forces with constant or
distributed strengths. A new dynamics of the collective rhythms is observed,
consisting of the external system superimposed on the autonomous one, a
characteristic feature of many thermodynamically open systems. This
deterministic, stable, continuously time-dependent, collective behaviour is
fully described, and the external impact to the original system is defined in both, the adiabatic and
non-adiabatic limits.

\end{abstract}


\maketitle

\section{Introduction}
\label{sec:Intro}

Biological examples provided the original motivation lying behind the Kuramoto
model (KM) of coupled phase oscillators  \cite{Strogatz_sync}. However, neither
the original model \cite{Kuramoto_book}, nor any of its extensions
\cite{Acebron_review}, have incorporated a fundamental property of living
systems -- their inherent time-variability. Many important characteristics of
open systems can be missed by not accounting for the non-equilibrium dynamics
that stems from their time-dependent (TD) parameters. Additionally, the
application of the KM to many problems would move closer to reality by allowing
for the natural frequency of each oscillator, or the coupling strengths, to be
externally modulated by TD forcing, as commonly occurs in living systems.
Among
the numerous collective rhythms traceable back to TD parameters, are frequency
flows in brain signals \cite{Rudrauf}, the modeling of brain dynamics under
anaesthesia \cite{Jane1} where anaesthetic strength modulates natural
frequencies \cite{Musizza}, event-related oscillatory responses of the brain
\cite{Pfurtschelle:99}, and the dynamics of cardiovascular ageing
\cite{yuri10}.
None of these are adequately described by existing models.
Additionally, similar considerations are to be expected in non-biological examples such as pattern formation
in a nonlinear medium far from equilibrium \cite{Lee2011}. Here, using trapped ions, one can vary the parameters at will and see the effect on the synchronization.

There has already been much work on coupled oscillators influenced by noise as
a special form of external dynamics \cite{Strogatz2}.
Likewise, driving by an external periodic force \cite{Shinomoto} is a long-explored model, characterized by the interplay to the phases of each oscillator between the external pacemaker and the mean field of all other oscillators.
A generalization of the KM that allowed certain time-varying frequencies and couplings was also numerically explored in \cite{Cumin2007}. However, the simulations were performed over a very small number of oscillators, the dynamics were not described analytically, and a qualitative description was not given for slow or fast varying cases.
Other studies of
non-constant collective rhythms include asymmetrically-coupled ensembles
\cite{Montbrio} and populations with multimodally distributed natural frequencies \cite{Bonilla1992}, with their complex mean field being a result of multimodal distribution of the parameters.
Frequency adaptation as discussed in \cite{Taylor2010} assumes non-constant natural frequencies, but without external influence. It is similar to the models with inertia \cite{Acebron_inertia} and its dynamics, apart from the stable incoherence, are characterized by either synchronization or bistable regime of both synchronized and incoherent states.
In addition, the model with drifting frequencies \cite{Rougemont} assumes frequency dynamics formulated as an Ornstein-Uhlenbeck process, but it also leads to time-independent mean fields, resembling the simple KM under influence of colored noise.
Alternately-switching connectivity \cite{So2008} or periodic couplings \cite{Lee2012}, are some of examples with varying coupling strengths. Yet, most of the discussions in these are concerned with the networks and graph theory properties of the system,
and only Heaviside step functions are considered for the interaction between oscillators.

Nevertheless, the TD mean
fields in most of these models result either from multistability or from unstable
equilibria.
Despite this, even in the cases where it stems from some external system \cite{Cumin2007, So2008, Lee2012}, the low-dimensional mean field dynamics and slow/fast reduced approaches are still missing.
As such, none of these models can
fully demonstrate the deterministic and stable
TD dynamics of many real physical, chemical, biological, or social systems that
can never be completely isolated from their surroundings. These systems do not
reach equilibrium but, instead, exhibit complex dynamical behavior that includes
the TD frequencies and couplings. We will show that our generalization of the
Kuramoto model encompasses these dynamics.

\section{Model}
\label{sec:Model}

An external, explicitly time-dependent, bounded function $x(t)$ is introduced.
It modulates the frequencies or couplings of the original model. This external influence
can also originate from another \cite{Jane:11} non-constant mean field. In the most
general case, the strengths of the interactions $I_i$ are distributed according
to a probability density function (PDF) $h(I)$, and likewise the distribution
$g(\omega)$ of the natural frequencies $\omega_i$. Thus, depending on which
parameter is influenced two generalized Kuramoto models emerge
 \begin{eqnarray}
 &A&: \ \ \ \dot{\theta}_i = \omega_i + I_i x(t)
    + K \ r(t) \sin[\psi(t)- \theta_i],  \label{eqn:NA_modelA} \\
 &B&: \ \ \ \dot{\theta}_i =\omega_i + [K + I_i x(t)] \ r(t) \sin[\psi(t) - \theta_i].
 \label{eqn:NA_modelB}
\end{eqnarray}
Here, a TD complex order parameter is introduced
\begin{eqnarray}
z(t)=r(t) \mathrm{e}^{i\psi(t)}=\frac{1}{N}\sum_{j=1}^{N} \mathrm{e}^{i\theta_j},
 \label{eqn:z}
\end{eqnarray}
where $r(t)$ and $\psi(t)$ are the TD mean-field amplitude and phase respectively.
For clarity, their explicit time-dependence will henceforth be omitted.

For each oscillator at any given time there is 1:1
correspondence between the fixed and TD parameters, i.e. \
$$\tilde{\omega}_i(t)=\omega_i+I_i x(t)$$ for model A, and
$$\tilde{K}_i(t)=K+I_i x(t)$$ for model B, or in general $$\tilde{I}_i(t)=I_i x(t).$$
Thus, for known  forcing $x(t)$, a single oscillator from both NA models can be uniquely defined by fixed parameters $\omega_i$ and $I_i$, or by the TD natural frequencies for the model A and TD couplings  for model B, $\tilde{\omega}_i$  and $\tilde{K}_i$ respectively, which in this case also encompass $x(t)$.
Similarly, instead of $\tilde{\omega}_i$ and $\tilde{K}_i$, $\omega_i$ and $\tilde{I}_i$  can be used, whereas  distributions of these TD variables accordingly  become $\tilde{g}(\tilde{\omega})$, $\tilde{\Gamma}(\tilde{K})$ and $\tilde{h}(\tilde{I})$.

To analyze the models (\ref{eqn:NA_modelA}), (\ref{eqn:NA_modelB})
the thermodynamic limit $N\rightarrow \infty$ is assumed.
Here, the state of the system with fixed forcing ($x(t)=const.$) would have been described by a continuous PDF $\rho(\theta,\omega,I,t)$ which gives the proportion of oscillators with phase $\theta$ at time $t$, for fixed $\omega$ and $I$ \cite{Mirollo2007}.
On the other hand, the one to one correspondence between the fixed and TD parameters in terms of PDFs implies that the same number of
oscillators can be described by either of the following PDFs
\begin{eqnarray}
\label{eqn:h1}
 |h(I)dI|=|\tilde{h} (\tilde{I}(I,t))\ d\tilde{I}|,
\end{eqnarray}
or
$$|g(\omega)d\omega| = |\tilde{g}(\tilde{\omega}(\omega,I,t))d\tilde{\omega}|$$ and $$|\Gamma(K,I)dK| = |\tilde{\Gamma}(\tilde{K}(K,I,t))d\tilde{K}|$$ if $\tilde{\omega}$ and $\tilde{K}$ are used for describing the population.
Also, the infinitesimal number of oscillators $d N$  is given by
\begin{eqnarray}
\label{eqn:rho1}
d N &=& |\rho(\theta,\omega,I,t) \ g(\omega) \ h(I) \ d \theta \ d\omega \ dI| = \nonumber \\
&& |\tilde{\rho}(\theta,\omega,\tilde{I},t) \ g(\omega) \ \tilde{h}(\tilde{I}) \ d \theta \ d\omega \ d\tilde{I}|,
\end{eqnarray}
where  PDFs $\rho$ and $\tilde{\rho}$  give the proportion of oscillators with
phase $\theta$ at time $t$, for given fixed $\omega$ and $I$, or fixed $\omega$ and TD $\tilde{I}$ respectively.
From probability theory it is known that by definition any PDF is nonnegative, and by substituting (\ref{eqn:h1}) into (\ref{eqn:rho1})  directly follows
\begin{eqnarray}
\label{eqn:rho2}
\rho(\theta,\omega,I,t)  = \tilde{\rho}(\theta,\omega,\tilde{I},t), \ \ \text{where $\tilde{I}=I x(t)$ }.
\end{eqnarray}
Analogously, for $\tilde{\omega}$ and $\tilde{K}$ instead of $\tilde{I}$, one would obtain $$\rho(\theta,\omega,I,t) =
\tilde{\rho}_1(\theta,\tilde{\omega},K,t) =
\tilde{\rho}_2(\theta,\omega,\tilde{K},t),$$ with $\tilde{\omega}=\omega+I x(t)$ and $\tilde{K}=K+I x(t)$.

Thereafter, the state of the oscillatory
system can be described either by a continuous PDF
$\rho(\theta,\omega,I,t)$ which assumes fixed parameters, or by its counterpart
$\tilde{\rho}(\theta,\omega,\tilde{I},t)$ with TD parameters.
However, since using PDF with TD parameters  would further complicate the continuity equation for fixed volume by including gradients along the TD variables also,
we choose to define the distribution for the fixed $\omega$ and $I$.
In this way, the only gradient of the PDF $\rho$ is along the phases.

The chosen probability density function $\rho$ is then normalized as $$\int_{0}^{2 \pi}
\rho(\theta,\omega,I,t)d\theta=1.$$
\noindent Moreover in the $\theta, \omega, I$ parameter space
the number of oscillators given by $\rho(\theta,\omega,I,t) g(\omega) h(I) d \theta d\omega dI$ for each natural frequency $\omega$ and strength $I$ of the forcing $x(t)$ is conserved, and only phases $\theta$ change with time.
Thus, the gradient along $\theta$ will be solely responsible for divergence of the oscillators.
Hence the continuity equation for every fixed $\omega$ and $I$ is given by
\begin{eqnarray}
\label{eqn:continuityA}
&A&: \ \ \frac{\partial\rho}{\partial t}=-\frac{\partial}{\partial \theta}\{[\omega + I x(t)+\frac{K}{2i}(z\mathrm{e}^{-i\theta}-z^\ast \mathrm{e}^{i\theta})]\rho\}, \  \\
&B&: \ \ \frac{\partial\rho}{\partial t}=-\frac{\partial}{\partial \theta}\{[\omega +\frac{K + I x(t)}{2i}(z\mathrm{e}^{-i\theta}-z^\ast \mathrm{e}^{i\theta})]\rho\}, \ \label{eqn:continuityB}
\end{eqnarray}
where the velocity along $\theta$ is substituted from the governing equations (\ref{eqn:NA_modelA}, \ref{eqn:NA_modelB}).
The definition (\ref{eqn:z}) is also included in (\ref{eqn:continuityA}, \ref{eqn:continuityB}), rewritten using
$$\frac{1}{N}\sum_{j} \sin(\theta_j-\theta_i)={\rm Im}\{z
\mathrm{e}^{-i\theta_i} \},$$ so that it becomes
\begin{equation}
\label{eqn:z2}
z = \int_{0}^{2\pi}\int_{-\infty}^{\infty}\int_{-\infty}^{\infty}\rho (\omega, I, \theta, t)  g(\omega) h(I) \mathrm{e}^{i\theta} d\theta d\omega d I.
\end{equation}

The same reasoning for preserving the number of oscillators would also apply for  $\tilde{\rho}(\theta,\omega,\tilde{I},t) g(\omega) \tilde{h}(\tilde{I}) d \theta d\omega d\tilde{I}$ if the infinitesimal volume of the  space $\theta, \omega, \tilde{I}$
is moving with $x(t)$ along the axis of the TD parameter, which in this case is $\tilde{I}$.
Thus again the only gradient of $\tilde{\rho}$ would be along phases, and continuity equations would have the same form as (\ref{eqn:continuityA}, \ref{eqn:continuityB}) with $I x(t)$ substituted with $\tilde{I}$, and $\rho$ with $\tilde{\rho}$.

\section{Low-dimensional dynamics}
\label{sec:OA}

Since $\rho (\theta,\omega,I,t)$ is real and $2\pi$ periodic in $\theta$, it
allows a Fourier expansion. The same would also hold for $\tilde{\rho}(\theta,\omega,\tilde{I},t) $.
Next, we apply  the Ott and Antonsen ansatz \cite{Ott1} in its coefficients, such that $$f_n(\omega,I,t)=[\alpha(\omega,I,t)]^n.$$ Thus,
\begin{eqnarray}
\label{eqn:ansatz}
\rho(\theta,\omega,I,t)=\frac{1}{2\pi}\{1+\{\sum_{n=1}^{\infty}{[\alpha(\omega,I,t)]^n\mathrm{e}^{in\theta}+{\rm c.c.}}\}\}, \ \ \ \
\end{eqnarray}
where c.c. is the complex conjugate.
Substituting (\ref{eqn:ansatz}) into the continuity equations (\ref{eqn:continuityA}, \ref{eqn:continuityB}), it follows that this special form of $\rho$ is their particular solution as long as $\alpha(\omega,I,t)$ evolves with
\begin{eqnarray}
&A&: \ \ \ \frac{\partial\alpha}{\partial t}+i[\omega+Ix(t)]\alpha+\frac{K}{2}(z\alpha^2-z^\ast)=0, \label{eqn:ansatz2A} \\
&B&: \ \ \ \frac{\partial\alpha}{\partial t}+i\omega\alpha+\frac{K+Ix(t)}{2}(z\alpha^2-z^\ast)=0,
\label{eqn:ansatz2B}
\end{eqnarray}
for models A and B respectively.
The same ansatz implemented in Eq.~(\ref{eqn:z2}), reduces the order parameter to
\begin{eqnarray}
\label{eqn:z*}
z^\ast=\int_{-\infty}^{+\infty}\int_{-\infty}^{+\infty} \alpha(\omega, I, t) g(\omega) h(I) d\omega dI.
\end{eqnarray}

Eqs.~(\ref{eqn:ansatz2A}, \ref{eqn:ansatz2B}) hold for any distributions of $\omega$ and $I$, and for any forcing $x(t)$. They describe the evolution of the parameter $\alpha$ which is related to the complex mean field through the integral equation (\ref{eqn:z*}).
These integrals can be analytically solved for certain distributions $g(\omega)$ and $h(I)$, thus directly leading to the low-dimensional evolution of $z$.
Hereafter we focus on all such cases and therefore in all further analysis the natural frequencies follow a Lorentizan
distribution, and $\alpha(\omega,I,t)$ is continued to the complex
$\omega$-plane so $g(\omega)$ can be written as $$g(\omega)=\frac{1}{2 \pi
i}[\frac{1}{\omega-(\hat{\omega}-i \gamma)}-\frac{1}{\omega-(\hat{\omega}+i
\gamma) }]$$ with poles $\omega_{p1,2}=(\hat{\omega} \pm i\gamma)$, where $\hat{\omega}$ is the mean of $g(\omega)$.

\subsection{Time-dependent natural frequencies}
\label{subsec:A}

The simplest
case of model A, Eq.~(\ref{eqn:NA_modelA}), is when the external forcing is
identical for each oscillator, $h(I)=\delta(I-\epsilon)$. This leads to trivial
dynamics, solved by simply making the reference frame rotate at the TD
frequency $\hat{\omega} + f(t)$, where $\hat{\omega}$ is the mean of
$g(\omega)$ and $f+\dot{f}t=\epsilon x$.

The non-autonomous (NA) dynamics arises for nonidentical forcing. We first
assume strengths proportional to frequencies, i.e.\
$$\tilde{\omega}(t)=\omega[1+\epsilon x(t)]$$ with a constant $\epsilon$.
This means that $I=\epsilon \omega$ and $h(I)=g(\epsilon \omega)$, and since  $\omega$ and $I$ in this case are not independent variables, the latter can be omitted in the PDF $\rho$.
Hence, the
integration in Eq.~(\ref{eqn:z*}) is now only over $\omega$, and by closing the
integral in any of the complex half-planes, is given by the residue of the
encircled pole. As a requirement from \cite{Ott1},
$|\alpha(\omega,t)|\rightarrow0$ as $\Im(\omega)\rightarrow\mp\infty$,
depending on which pole is encircled. The last limit transforms
Eq.~(\ref{eqn:ansatz2A}) into $\frac{\partial \alpha}{\partial t} =  -
\tilde{\omega}(t) \alpha$. Thus, for $[1+\epsilon x(t)]>0$, the encircling is
around the pole $\omega_{p2}=(\hat{\omega}-i\gamma)$, while for $[1+\epsilon
x(t)]<0$ the upper half-plane encircling involves
$\omega_{p1}=(\hat{\omega}+i\gamma)$. Next, the residue at these poles,
$$z^\ast=\alpha(\hat{\omega}\mp i\gamma,t ),$$ is substituted in
Eq.~(\ref{eqn:ansatz2A}) yielding
\begin{eqnarray}
\label{eqn:evolution_z1}
\dot{r}=-r[\gamma |1+\epsilon x(t)| +\frac{K}{2}(r^2-1)]; \
\dot{\psi}=\hat{\omega}[1+\epsilon x(t)]. \ \ \ \ \
\end{eqnarray}
The ansatz (\ref{eqn:ansatz}) holds only for nonidentical oscillators
\cite{Ott2}, implying the requirement $\tilde{\omega}(t)\neq0, \forall t$.

If the previously discussed alternative continuity equation for $\tilde{\rho}$ was used, then $\alpha(\omega,I,t)$ would become $\tilde{\alpha}(\omega,\tilde{I},t)$ and the the poles of $\tilde{I}$ would be TD.
Nevertheless,  substituting  $\tilde{\alpha}(\omega,{I},t)$  into the continuity equation that includes $\tilde{I}$ would  lead to the same evolution for the mean field, thus confirming the analysis.

\begin{figure}[t!h]
\centering
\includegraphics{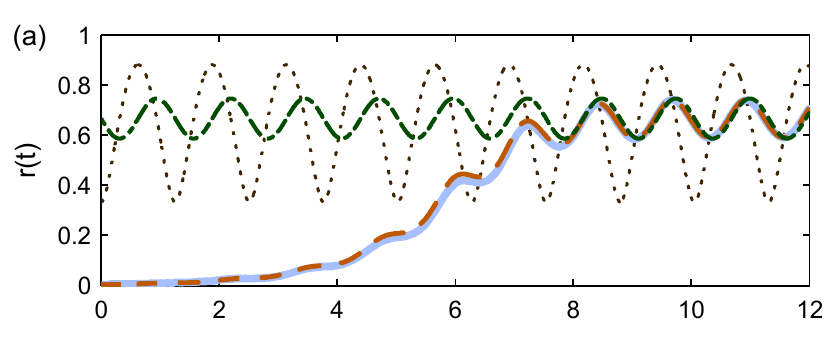}\\
\centering
\includegraphics{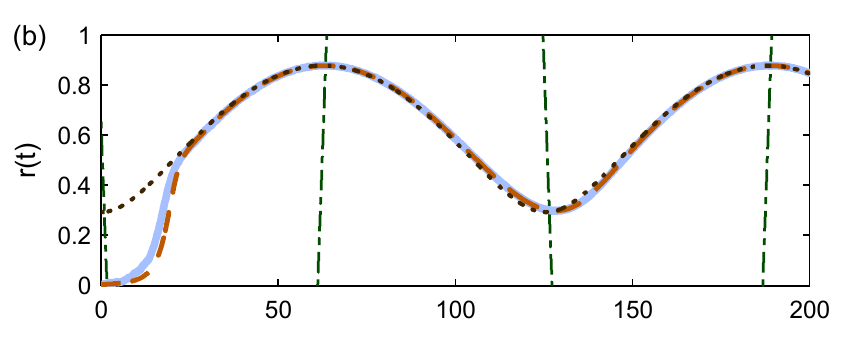}
\centering
\includegraphics{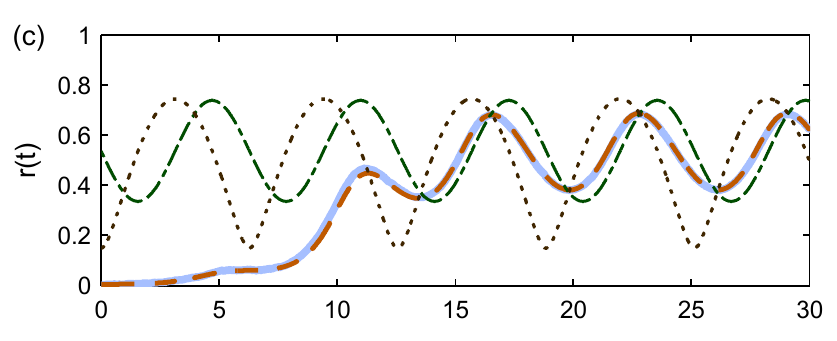}\\
\centering
\includegraphics{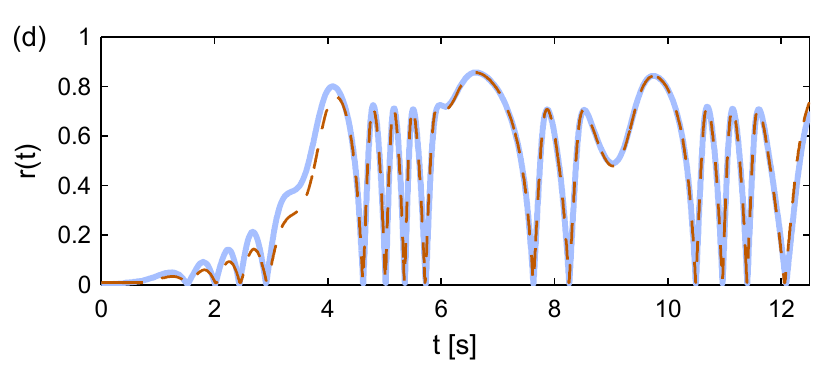}
\caption{(color online) The time-varying mean field for model A, Eq.\ (\ref{eqn:NA_modelA})
resembles the externally applied cosine (a-c), or  chaotic forcing (d). Numerical
simulations of the full system  Eq.\ (\ref{eqn:NA_modelA}) (light blue) are in
agreement with the low-dimensional dynamics (dashed red):
Eqs.\ (\ref{eqn:evolution_z1}-\ref{eqn:evolution_z5}) (see text for details). Adiabatic (dotted brown),
 and
non-adiabatic evolutions (dashed-dotted green),
Eqs.~(\ref{eqn:r_adiabatic}-\ref{eqn:r_non-adiabatic3}), confirm the reduced dynamics in its limits (see text for details). 
The distribution  $h(I)$
is: (a-b) same as $g(\omega)$, $K=3.5, \epsilon=0.6$, $\Omega=5$ and
$\Omega=0.05$ respectively; (c) independent Lorentzian, $K=4.5$, $\gamma_I=0.6$ and $\Omega=1 $; 
and (d) bimodal $\delta$, $K=8$, $\gamma=1$, $\gamma_I=1$ and $\hat{I}=1$.
}
\label{fig:a}
\end{figure}
Model A is also solvable with an independent Lorentzian distribution of forcing
strengths. The frequencies follow $\tilde{\omega}(t)=\omega+Ix(t)$ and the mean
and half-width of $h(I)$ are $\hat{I}$ and $\gamma_I$ respectively. The
integrals in Eq.~(\ref{eqn:z*}) can again be closed in the lower or upper
complex half-plane, and the requirements for $\alpha(\omega,I,t)$ are similar
to those in the previous case. Hence, the $I$ integral for $x(t)>0$ is around
the pole $I_{p1}=(\hat{I}+i\gamma_I)$ and around $I_{p2}=(\hat{I}-i\gamma_I)$
otherwise, while in the $\omega$ integral the encircling is around the pole
$\omega_{p2}=\hat{\omega}-i\gamma$. Thus, the residues give $$z^\ast =
\alpha(\hat{\omega}-i\gamma,\hat{I}-i\gamma_I,t),$$ which is applied in
Eq.~(\ref{eqn:ansatz2A}), so we finally obtain
\begin{eqnarray}
\label{eqn:evolution_z3}
\dot{r}=-r[\gamma + \gamma_I |x(t)| +\frac{K}{2}(r^2 - 1)], \ \dot{\psi}=\hat{\omega}  + \hat{I} x(t). \ \
\end{eqnarray}
A similar analysis would be possible for any other polynomial Lorentzian-like distributions of
$\omega$ and $I$.

The only other analytically solvable form of model A that we are aware of is
with multimodal $\delta$-distributed external strengths. For simplicity we
choose the bimodal function $$h(I)=\frac{1}{2}[\delta(I-\hat{I}-\gamma_I) +
\delta(I-\hat{I}+\gamma_I)].$$ The integral (\ref{eqn:z*}) now leads to
\begin{eqnarray}
\label{eqn:evolution_z51}
 z^\ast=\frac{1}{2}[\alpha_1(\hat{\omega}-i\gamma,\hat{I}-\gamma_I,t)+\alpha_2(\hat{\omega}-i\gamma,\hat{I}+\gamma_I,t)], \ \ \ \ \
\end{eqnarray}
with dynamics consistently described by the evolutions of $\alpha_{1,2}$
obtained from Eq.~(\ref{eqn:ansatz2A}),
\begin{eqnarray}
\label{eqn:evolution_z5}
 && \frac{\partial \alpha_{1,2}}{\partial t} = -\{i[\hat{\omega}+(\hat{I} \mp \gamma_I)x(t)]-\gamma\}\alpha_{1,2} +  \nonumber\\
 && \ \ \ \ \ \ \ \ \ \ + \frac{K}{4} [\alpha_{1}+\alpha_{2} - \alpha_{1,2}^2(\alpha_{1}+\alpha_{2})^\ast].
\end{eqnarray}
This case of model A was also investigated in \cite{Choi}, where Choi \emph{et al.}  carried out a
bifurcation analysis near the limit $r K \ll 1$.

Following the restrictions on $x(t)$ in the problems analyzed in Fig.\
\ref{fig:a}, we took $$x(t)= \cos \Omega t \ \ \text{and} \ \ \epsilon<1$$ in the case
of strength proportional to the frequency, while for model A with independent
Lorentzianly-distributed strengths, the forcing is $$x(t)= 1+ \cos \Omega t.$$
Finally, for bimodal $\delta$-distributed  strengths, the absence
of restrictions on the external field allows it to be the $x$ component of a
R\"{o}ssler oscillator \cite{Rossler}. In all the problems shown, the NA TD
dynamics is revealed and fully described by the reduced NA low-dimensional
system. A Runge-Kutta 4 algorithm was used for numerical integration of
Eqs.~(\ref{eqn:NA_modelA}, \ref{eqn:NA_modelB}) over $100000$ oscillators, with
a time-step of $0.0025 s$, while half-width and mean of the natural frequencies
were $\gamma=1$ and $\hat{\omega}=0$, except where otherwise stated.

\subsection{Time-dependent coupling strengths}
\label{subsec:B}

We have also investigated the low-dimensional evolution of NA model B,
Eq.~(\ref{eqn:NA_modelB}). Since all the couplings in the original model are
equal, there is no qualitative difference between the situations with identical
forcing to each coupling, and coupling-dependent forcing.  We chose the
latter and proceed as for model A, yielding
\begin{eqnarray}
\label{eqn:evolution_z2}
\dot{r}=-r[\gamma+\frac{K}{2}[1+\epsilon x(t)] (r^2 - 1) ];  \ \ \
\dot{\psi} =  \hat{\omega}. \ \ \ \
\end{eqnarray}
The analysis for multimodal $\delta$-distributed strengths is also very
similar to that for model A (\ref{eqn:NA_modelA}). E.g. for bimodal $h(I)$,
Eq.~(\ref{eqn:evolution_z51}) holds again with $\alpha_{1,2}$ evolving as
\begin{eqnarray}
\label{eqn:evolution_z6}
 && \frac{\partial \alpha_{1,2}}{\partial t} = -(i\hat{\omega}-\gamma)\alpha_{1,2} + \frac{1}{4} K [1 + (\hat{I} \mp \gamma_I) x(t)] \nonumber\\
 && \ \ \ \ \ \ \ \ \ \ \times [\alpha_{1}+\alpha_{2} - \alpha_{1,2}^2(\alpha_{1}+\alpha_{2})^\ast].
\end{eqnarray}
\begin{figure}[t!h]
\centering
\includegraphics{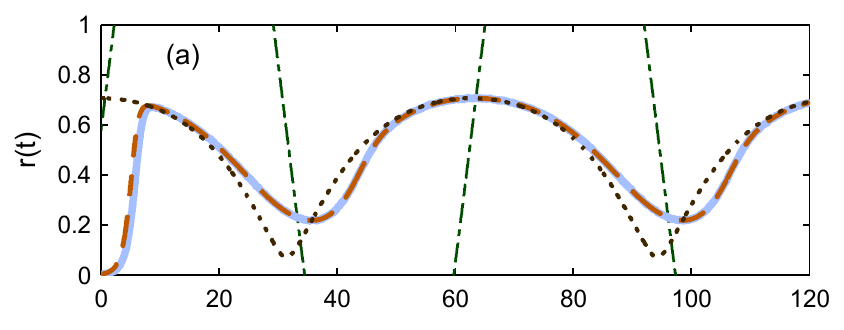}\\
\centering
\includegraphics{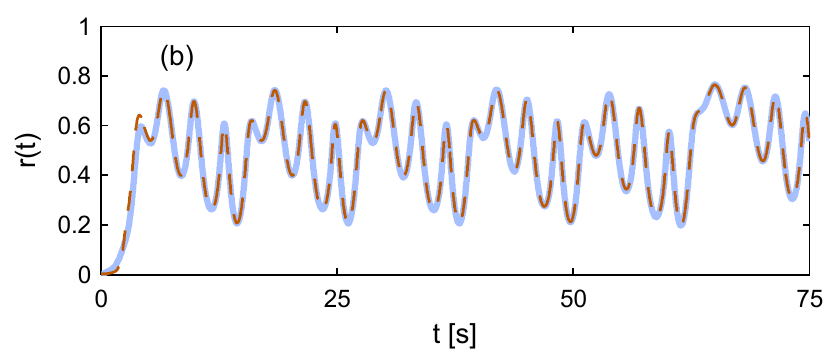}
\caption{(color online)  Time-varying mean field for model B, Eq.~(\ref{eqn:NA_modelB}),
follows the external cosine (a) or the chaotic (b) forcing. Numerical
simulation of Eq.\ (\ref{eqn:NA_modelB}) (light blue) coincides with the
low-dimensional evolution (dashed red), Eq.~(\ref{eqn:evolution_z2}) -- (a) and
Eq.~(\ref{eqn:evolution_z6}) -- (b). (a) Adiabatic (dotted brown),
Eq.~(\ref{eqn:delta27}), and non-adiabatic evolution (dashed-dotted green),
Eq.~(\ref{eqn:r_non-adiabatic2}) for constant forcing with $K=3$, $\Omega=0.1$
and $\epsilon=0.33$. (b) Bimodal $\delta$-distributed strengths with $K=5$,
$\gamma_I=1$ and $\hat{I}=0$. }
 \label{fig:b}
\end{figure}
However, for a Lorentzian distribution $h(I)$, contour integration cannot be applied to  Eq.~(\ref{eqn:z*}).
Namely, the integration contour should be such that if $\alpha(\omega, I, t)$ is analytic and $|\alpha|\leq1$ everywhere inside the contour at $t=0$, this would also hold for all $t>0$.
However,  for this to happen, one of the  requirements from \cite{Montbrio2011}  is
$|\alpha| \leq 0$, for $|\alpha|=1$. This should be taken in regard to the semicircular integration path $I=|I|e^{i \vartheta}$ with $|q|\rightarrow\infty$ and $\vartheta\in(0, \pi)$ or $\vartheta\in(-\pi, 0)$ depending on the half-plane of the contour.
Thus, substituting for $I$ into Eq.~(\ref{eqn:ansatz2B}) and taking $|\alpha|=1$ , it yields
\begin{eqnarray}
\label{eqn:conditionB_Lor}
\frac{\partial |\alpha|}{\partial t}=|I|x(t) r \sin\vartheta \sin [\phi(\omega, I, t)-\psi(t)].
\end{eqnarray}
Here, $\phi$ is the phase of $\alpha$ that depends on $\omega$, $I$ and $t$, implying that the last sine can have either signs. Consequently,  it cannot be proven that the condition $\frac{\partial {|\alpha|}}{\partial t} \leq 0$  holds for $\forall$ $t$ and $\omega$, on either of the half-planes. As a result, the integral in the Eq.~(9) cannot be solved for $I$ using the residue theorem.

In contrast,
the restrictions do not affect the NA parts of the other discussed variations
of model B. To confirm this generality, $x(t)$ for the problem shown in
Fig.~\ref{fig:b}~(b) is a chaotic signal from a R\"{o}ssler oscillator.
Similarly, the chosen amplitude of the cosine forcing in Fig.~\ref{fig:b}~(a)
allows close-to-incoherent dynamics to be observed in some intervals, so that appear
the limitations of the slow-fast approaches discussed in the following section.

A theorem in  \cite{Ott2} states that  Eqs.~(\ref{eqn:ansatz2A},
\ref{eqn:ansatz2B}) asymptotically capture all macroscopic behavior of the
system as $t\rightarrow\infty$. Moreover the incoherent and partly synchronized
states both belong to the manifold defined by Eqs.~(\ref{eqn:ansatz2A},
\ref{eqn:ansatz2B}) \cite{Ott1}, and initial incoherent state is set with
uniformly distributed phases at time $t=0$. Thereafter, the ansatz
(\ref{eqn:ansatz2A}, \ref{eqn:ansatz2B}) and the evolutions
(\ref{eqn:evolution_z1}-\ref{eqn:evolution_z6}) should continuously describe our system, as confirmed by Figs.~\ref{fig:a} and \ref{fig:b}.

\section{Reduced dynamics}
\label{sec:RD}
The plots in Figs.~\ref{fig:a}~(a)-(c) and
\ref{fig:b}~(a) show that the oscillations of the mean field follow the frequency of the external forcing,
but this raises the questions of what is the amplitude of the oscillations and whether they can adiabatically follow the forcing.
Similarly, an obvious feature of the same results  is the low-frequency filtering of the external fields, i.e.\
the only difference between plots (a) and (b) of Fig.~\ref{fig:a} is the
frequency of the external forcing, while its influence is much more prominent
in the latter. This is actually a well-known, but not much explored,
characteristic of population models, and it is a direct consequence of their
intrinsic transient dynamics \cite{Strogatz_sync}.

In the following we adopt
fast-slow reduction to simplify the evolution for simple periodic forcing. The
reduction depends on the period of the external field $T=2\pi/\Omega$, relative
to the system's transition time, $\tau$, and has not been applied to similar
systems. The exponential damping rate of the original system is defined by $\tau$ \cite{Ott1} and
$$\tau=1/|K/2-\gamma|.$$ For a system far from
incoherence, $K=2\gamma+{\rm O}(2\gamma)$, $\tau\approx1/{\rm O}(\gamma)$
holds, meaning that the transition time depends only on the width of the
distribution of natural frequencies. Thereafter for this case, the system's
response is adiabatic for slow external fields, $\Omega\ll\gamma$, and
non-adiabatic for fast, $\Omega\gg\gamma$. From now on, the dependence on $\gamma$ is
removed by scaling the time and the couplings in the autonomous system,
$t=t/\gamma$, $K=K/2 \gamma$ and $\tau=1/|K-1|$ (the scaled variables keep the
same letters).

For model A, Eq.~(\ref{eqn:NA_modelA}), with $x(t)= \cos \Omega t$ and $h(I)=\delta(I-\epsilon)$,
after the initial transition and in the absence of bifurcations, the amplitude of
the mean field consists of a constant term $r_0$ and a TD term $\Delta r (t)$.
For the non-adiabatic response, simulations, grey lines in Fig. \ref{fig:a}(a),
show that $\Delta r(t)\sim 1/\Omega$ and $r_0 \gg\Delta r(t)$. Thereafter $r_0$
can be expressed as averaged over one period  $T=2\pi/\Omega$ of the
oscillations of $\Delta r(t)$. This way it follows $0=-r_0+K(r_0^3-r_0)$ \cite{comment}, or
$$r_0=\sqrt{1- 1/K}.$$ Further, we apply $r(t)\approx r_0$ and $\frac{d
r}{dt}=\frac{d \Delta r}{dt}$ to Eq.~(\ref{eqn:evolution_z1}) and then
integrate it. From there  $$\Delta r(t)=-r_0\frac{\epsilon}{\Omega}\sin \Omega
t,$$ and the magnitude of the NA response is
\begin{eqnarray}
\label{eqn:delta6}
\Delta_{\rm fast} = 2\frac{\epsilon}{\Omega} \sqrt{1- \frac{1}{K}}.
\end{eqnarray}
Hence the long-term non-adiabatic evolution follows
\begin{eqnarray}
\label{eqn:r_non-adiabatic}
r_{\rm fast}(t) = \sqrt{1- \frac{1}{K}} ( 1 -  \frac{\epsilon}{\Omega} \sin \Omega t).
\end{eqnarray}

\noindent The adiabatic behavior emerges through the introduction of a slow
time-scale $t'=\Omega t$, such that the system is constant on the fast
time-scale $t$, and changes only in $t'$. Hence the l.h.s. of
Eq.~(\ref{eqn:evolution_z1}) is zero, whence
\begin{eqnarray}
\label{eqn:r_adiabatic}
r_{\rm slow}(t) = \sqrt{1-\frac{1 + \epsilon \cos \Omega t}{K}},
\end{eqnarray}
while, for the magnitude of the NA part, we obtain
\begin{eqnarray}
\label{eqn:delta7}
\Delta_{\rm slow}= \sqrt{1-\frac{1 - \epsilon}{K}} - \sqrt{1-\frac{1 + \epsilon}{K}}.
\end{eqnarray}
An analogous analysis can be performed for the appropriate form of model B,
Eq.~(\ref{eqn:NA_modelB}), leading to the low-dimensional evolution for fast
cosine forcing given by
\begin{eqnarray}
\label{eqn:r_non-adiabatic2}
r_{\rm fast}(t) = \left( 1 +  \frac{\epsilon}{\Omega} \sin \Omega t\right) \sqrt{1- \frac{1}{K}},
\end{eqnarray}
and for slow forcing
\begin{eqnarray}
\label{eqn:delta27}
r_{\rm slow}(t) = \sqrt{1-\frac{1}{K (1 + \epsilon \cos \Omega t)}}.
\end{eqnarray}
For the dynamics of model A with independent Lorentzian strengths and cosine forcing,
$x(t)=1+ \cos \Omega t$, the time is scaled by $(\gamma + \gamma_I)$, and the
dynamics follows
\begin{eqnarray}
\label{eqn:r_non-adiabatic3}
r_{\rm fast}(t) = \sqrt{1- \frac{1}{K}} [ 1 -  \frac{\gamma_I}{\Omega(\gamma + \gamma_I)} \sin \Omega t]
\end{eqnarray}
for fast forcing, while for slow driving
\begin{eqnarray}
\label{eqn:r33}
r_{\rm slow}(t) = \sqrt{1-\frac{1}{K}-\frac{\gamma_I \cos \Omega t}{K (\gamma + \gamma_I)}}.
\end{eqnarray}
The adiabatic responses can also be obtained from the self-consistency of
Eqs.\ (\ref{eqn:continuityA}, \ref{eqn:z2}) for stationary states of the mean
field. Namely, assuming very slow dynamics of the external forcing, the system
can be treated as quasistationary. This is similar to assuming stationarity on
a fast time scale. Thus one obtains $$r=\sqrt{1- 2 \gamma(t) / K(t)},$$ which
corresponds to the results (\ref{eqn:r_adiabatic}), (\ref{eqn:delta27}).
\begin{figure}[t!]
\hspace{10pt} 
\includegraphics{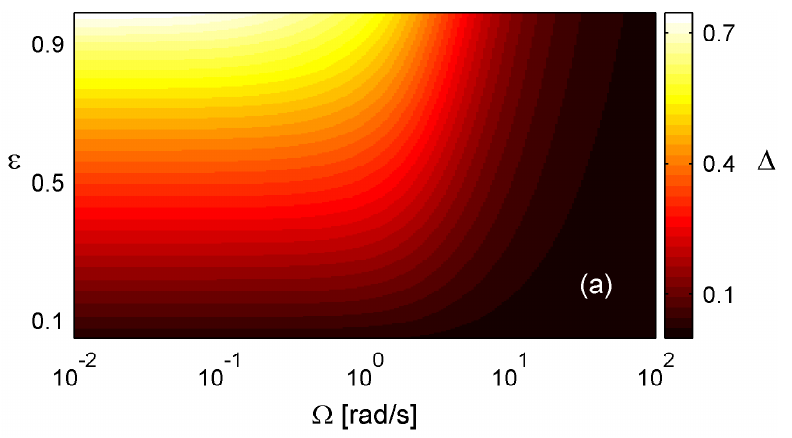}\\
\hspace{-20pt}
\includegraphics{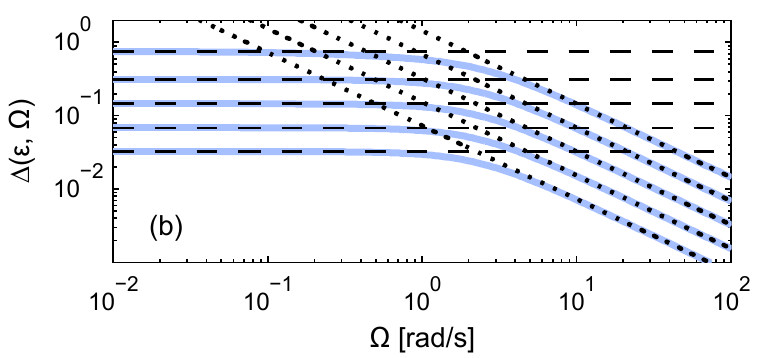}
\caption{ (color online) Magnitude of the response, $\Delta(\epsilon, \Omega)$, of the NA model A to the cosine forcing, Eq.~(\ref{eqn:NA_modelA}).
 External forcing strengths follow the distribution of frequencies, $K=4.25$ and $\Omega\in[10^{-2}, 10^2]$. (a) Results from  Eq.~(\ref{eqn:evolution_z1}) for  $\epsilon\in[0.05, 0.99]$. (b) Non-adiabatic (dotted black), Eq.~(\ref{eqn:delta6}) and adiabatic, Eq.~(\ref{eqn:delta7}), (dashed black) evolution for $\epsilon\in\{0.05, 0.1055, 0.2225, 0.4693, 0.99 \}$, compared with the real dynamics (light blue), Eq.\ (\ref{eqn:evolution_z1}).
}
 \label{fig:c}
\end{figure}

All the evolutions for reduced dynamics, Fig. \ref{fig:a}(a)-(c) and
\ref{fig:b}(a), are in line with the above analysis, confirming the interplay
between external and internal time scales of the NA system. The magnitudes of
the slow/fast responses to cosine forcing are given in Fig. \ref{fig:c} for
model A, Eq.~(\ref{eqn:NA_modelA}), with forcing strengths following the frequencies' distribution.
They  confirm the obtained dependance of
$\Delta$ on the frequency and amplitude of the external field. The
low-frequency filtering mentioned before is also obvious. The transient
behavior for slow and fast forcing can be seen in Fig. \ref{fig:c}(b), where
$\Delta$ is shown for both the actual and the reduced dynamics. This plot
perfectly matches the analytic limits for application of the reduction
approaches. Similar plots can also be obtained for the other problems analyzed.
However, for coupling  close to  critical, the system's transition time
increases and for $K \approx K_{c}$, $\tau \rightarrow \infty$. As a result,
the slow dynamics fails, as shown in Fig. \ref{fig:b}(a) at the minima of $r$ when it is close to
$0$, unlike the case $K=K_c+{\rm O}(K_c)$ given in Fig. \ref{fig:a}(a)-(b) or
Fig. \ref{fig:b}(a) for $r$ far from 0.

\section{Discussion}
\label{sec:D}

With the analysis of the reduced dynamics, supplementing the full
low-dimensional description, all aspects of the TD KM have been demonstrated.
The former is shown only for simple periodic forcing, but this does not
decrease the generality of the reduction, since any external field can be
represented by its Fourier components.
These methods are of great
importance in modeling systems with multiple time-scales of oscillation and
interaction, such as  the human cardiovascular system \cite{Aneta1}, or
inhibitory neurons in the cortex \cite{Wechselberger:09}.

In summary, we have characterised a new dynamics of interacting oscillators
subject to continuous, deterministic perturbation. It consists of the dynamics
of an external system superimposed on the original collective rhythm and was
missing from earlier models \cite{Acebron_review}, possibly leading to an incorrect
interpretation of some real dynamical systems. We have derived the impact of
the forcing and evaluated the effect of its dynamics, amplitude and
distribution. Thus, we have proposed a generalization of the Kuramoto model
that encompasses NA systems \cite{Rasmussen_book} and is directly applicable to
any thermodynamically open system. For example, the observed time-variations of brain
dynamics can be easily explained as a consequence of TD frequencies or
couplings of the single neurons, where the source of the external variation
could be due to anaesthesia \cite{Jane1}, event-related \cite{Pfurtschelle:99},
or due to some influence from another part of the brain. In particular, the
stable, time-varying mean field can now be reconstructed and, in this way, a
large range of systems tackled by the Kuramoto model -- spanning from a single
cell up to the level of brain dynamics -- can be described more realistically.

\section*{Acknowledgments}

We thank P. V. E. McClintock  and G. Lancaster for useful comments on the manuscript, and A. Duggento, L. Basnarkov, Y. Suprunenko and D. Iatsenko for valuable discussions.
The work was supported by the EPSRC (UK) and by a Lancaster University PhD grant.


\begin{thebibliography}{99}
\bibitem{comment} This is obtained from averaging both sides of Eq.~(\ref{eqn:evolution_z1}) for one period. The term $\Delta r(t)\cos \Omega t$ in the integral vanishes only if $\Delta r(t)\sim \sin \Omega t$, which is self-consistently proved as the obtained form of $\Delta r(t)$ for non-adiabatic response, Eq. (\ref{eqn:r_non-adiabatic}), follows this assumption.
\end{thebibliography}
\end{document}